\documentclass{article}
\usepackage{hiph-preprint}
\usepackage{amssymb}
\usepackage{epsfig}
\volnumber{22} \issuenumber{1} \edyear{2005}                             
\frompage{000} \topage{000}                                              
\recrevdate{\today}                                              
\def\rightmark{Di-jet correlation in Au + Au and Cu + Cu collisions from PHENIX}
\def\leftmark{J. Jia}

\title{Di-jet correlation in Au + Au and Cu + Cu collisions from PHENIX}
\authors{
{Jiangyong Jia
\index{One, A.} 
\index{Two, A.} 
}\\[2.812mm]
{\normalsize
\hspace*{-8pt}Columbia University and Nevis Laboratories, Irvington, NY 10533, USA\\[0.2ex]
}}

\abstract{PHENIX has measured the two particle azimuth correlation
in Au + Au at $\sqrt{s}$ = 200 GeV. Jet shape and yield at the
away side are found to be strongly modified at intermediate and
low $p_T$, and the modifications vary dramatically with $p_T$ and
centrality. At high $p_T$, away side jet peak reappears but the
yield is suppressed. Similar jet strength is found for Au + Au and
Cu + Cu collisions with similar number of participant nucleons.}


\makeindex
\begin{document}
 \maketitle
\section{Introduction}
We summarize the two particle azimuthal correlation results
presented in QM2005 poster. Due to space limitation, please refer
to ~\cite{Grau:2005sm,Jia:2005ab} for the technical details and
the results on reaction plane dependence of the jet correlation.
\section{Jet Correlation at Intermediate $p_T$ in Au + Au}
The Au + Au results are obtained from 1 billion minimum bias
events. According to~\cite{Jia:2005ab}, we parameterized the
correlation function $C(\Delta\phi)$ (CF) as,
\begin{eqnarray}
\label{eq:1} C(\Delta\phi) = J(\Delta\phi) + \xi\left(1+2v_2^{t}
v_2^{a} \cos\Delta 2\phi\right)
\end{eqnarray}
$J(\Delta\phi)$ represents the contribution from jet. $v_2^t$ and
$v_2^b$ are the elliptic flow for the trigger and associated
particles, respectively. $\xi$ is the only free normalization
factor, which is fixed using the ZYAM
assumption~\cite{Ajitanand:2005jj,Adler:2005ee}.

\begin{figure}[t]
\begin{center}
\epsfig{file=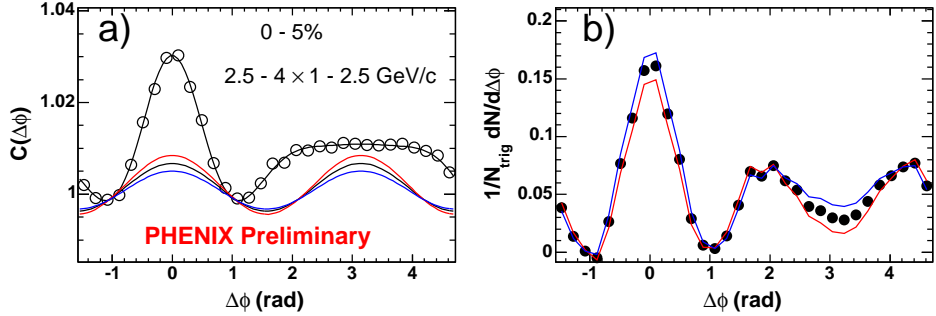,width=1\linewidth}
\caption{\label{fig:zyamcf1} a) Correlation function in 0-5\%
centrality bin, the lines indicated the level of flow background
and the systematic errors. b) Correspondingly background
subtracted per-trigger yield.}
\end{center}
\end{figure}

A typical correlation function from central Au + Au collisions is
shown in Fig.\ref{fig:zyamcf1}. $v_2$ values in the corresponding
$p_T$ selections are about 0.062 in 2.5-4 GeV/c, and 0.041 in
1-2.5 GeV/c ranges. The away side shape is very broad and
non-gauss like. It has a wide plateau that expands to about
$\pi\pm1$ radian. After the flow contribution (shown by the
curves) is subtracted, a dip appears around $\pi$ as shown in
Fig.\ref{fig:zyamcf1}b. This shape can not be due to the random
walk type of broadening of the jets from energy loss. It is
qualitatively consistent with Cherenkov gluons~\cite{cheren} or
shock wave~\cite{Casalderrey-Solana:2004qm} excited by the
travelling jets in the medium.

\begin{figure}[b]
\begin{center}
\epsfig{file=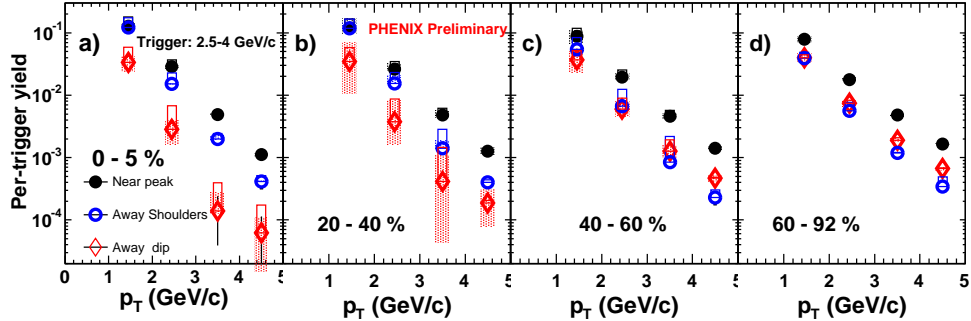,width=1\linewidth}
\caption{\label{fig:yield} The yield for trigger 2.5-4 GeV/$c$
plotted as function of associated hadron $p_T$ for four different
centrality bins.}
\end{center}
\end{figure}

To quantify the modifications of the jet shape, we study the jet
yield in three different $\Delta\phi$ regions: near side jet
region ($|\Delta\phi|<\pi/3$), the away side dip region
($|\Delta\phi-\pi|<\pi/6$), and the away side shoulder region
($|\Delta\phi-\pi\pm\pi/3|<\pi/6$). The shoulder region is
sensitive to the novel medium effects, while the dip region is
sensitive to the punch through jet contribution.
Fig.\ref{fig:yield} plots the jet yields in the three regions as
function of $p_{T,\rm{assoc}}$ for four centralities. In 0-5\%
centrality bin, there is a large separation between the yields for
the dip region and near side jet region, persistent to large
$p_T$. In more peripheral collisions, the yield of the dip region
becomes closer or even exceeds that for the shoulder region,
consistent with the returning of the away side jet to a normal
gauss shape.

\begin{figure}[t]
\begin{center}
\epsfig{file=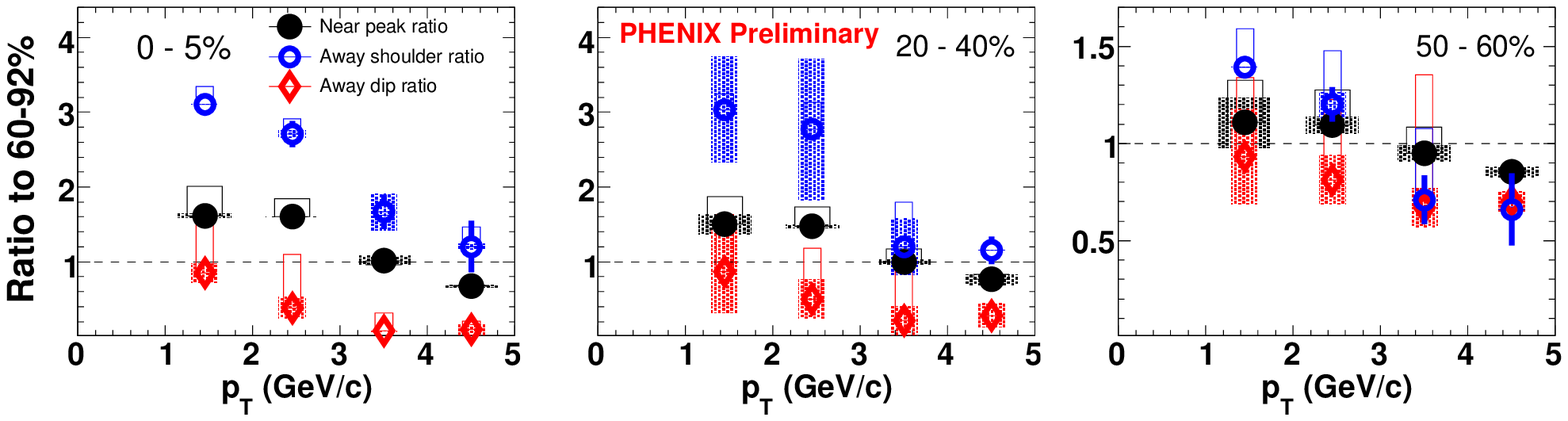,width=1\linewidth}
\caption{\label{fig:yield1} Ratio of the per-trigger yield between
central bin and 60-92\% centrality bin ($I_{\rm{cp}}$) for the
three $\Delta\phi$ ranges}
\end{center}
\end{figure}

The modification of the jet shape and yield can be quantified by
$I_{\rm{cp}}$, i.e. the ratios of the per-trigger yield between
central bin and 60-92\% peripheral bin, as shown in
Fig.\ref{fig:yield1} for the three $\Delta\phi$ regions. In 0-5\%
central bin, $I_{\rm{cp}}$ is well above one for the away shoulder
region at low $p_{T,\rm{assoc}}$, and decreases toward larger
$p_{T,\rm{assoc}}$. This is in sharp contrast with the away dip
region, which shows a suppression of the $I_{\rm{cp}}$, although
both regions have similar $p_T$ dependence. We also see a modest
enhancement of the per-trigger yield at the near side at low
$p_{T,\rm{assoc}}$. This plus the strong suppression in the away
side dip region are consistent with STAR's observation in a
earlier publication~\cite{Adler:2002tq} (for a somewhat different
$p_T$ selection). Finally, $I_{\rm{cp}}$ for all three
$\Delta\phi$ regions approach unity towards peripheral collisions.

\begin{figure}[t]
\begin{center}
\epsfig{file=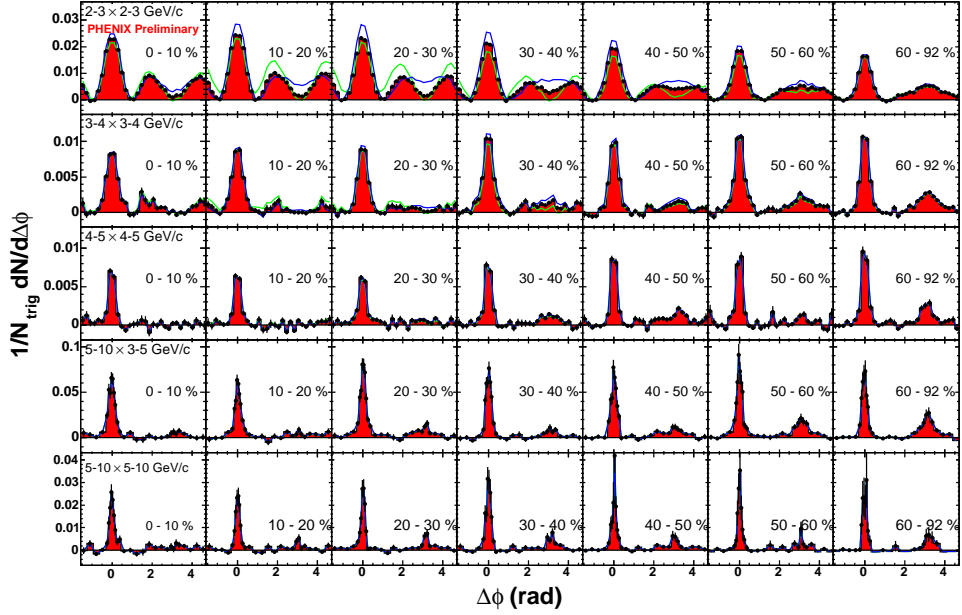,width=1\linewidth}
\caption{\label{fig:scan} Background subtracted per-trigger jet
yield in $\Delta\phi$ as function of $p_T$ (vertical) and
centrality (horizontal).}
\end{center}
\end{figure}
\section{Jet Correlation from Low $p_T$ to High $p_T$ in Au + Au}\label{maths}
Jet correlations at different $p_T$ reflect different aspect of
the interaction between jet and the medium. The results are
summarized in Fig.\ref{fig:scan}, where we plot the per-trigger
yield as function of both $p_T$ (vertical) and centrality
(horizontal). Along the vertical direction, we can see how the
away side jet evolves from a cone type of structure at
intermediate $p_T$ to a relatively flat distribution at moderately
high $p_T$, to a reappeared jet structure at high $p_T$. Along the
horizontal direction, we can also see that the modifications of
the away side jet shape depend strongly on centrality. Further
detailed discussions can be found in~\cite{Jia:2005ab}.

\section{Comparison of High $p_T$ $\pi^0-h$ Correlation Between Au + Au and Cu + Cu}
\begin{figure}[t]
\begin{center}
\epsfig{file=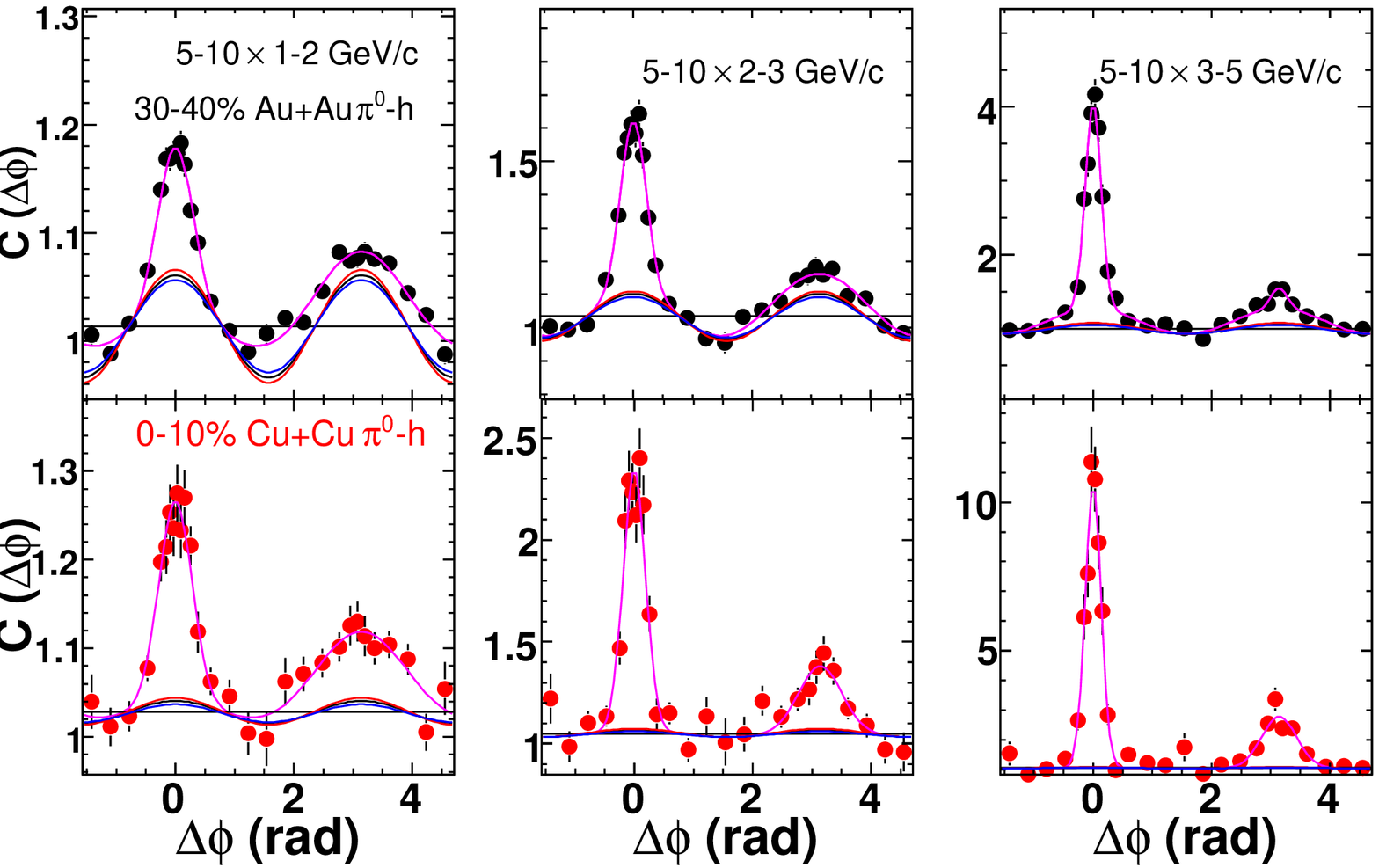,width=0.8\linewidth}
\caption{\label{fig:auaucucu} The $\pi^0$ -h correlation functions
for three different associated charged hadron $p_T$ ranges (with
fixed triggering $\pi^0$ $p_T$: 5-10 GeV/$c$). (Top row) 30-40\%
most central Au + Au centrality bin, and (Bottom row) 0-10\% most
central Cu + Cu centrality bin.}
\end{center}
\end{figure}
If jet modifications are mainly determined by the size of the
medium created in the heavy-ion collisions, we should expect a
similar modification in Cu + Cu collisions with similar
$N_{\rm{part}}$ as in Au + Au. However, they have very different
$v_2$ systematics due to their totally different shape of the
overlap region. Fig.\ref{fig:auaucucu} shows the comparison of the
high $p_T$ $\pi^0-h$ correlation functions together with the
estimated flow background for 30-40\% Au + Au centrality and
0-10\% Cu + Cu centrality bin. Both have similar number of
participants and number of collisions: $N_{\rm{part}}$ = 98 and
$N_{\rm{coll}}$ = 183 in Cu + Cu and $N_{\rm{part}}$ = 114 and
$N_{\rm{coll}}$ = 220 in Au + Au. The amplitudes of the
correlation functions, which reflect the ratio of jet signal to
the combinatoric background, are larger in Cu + Cu case because Cu
+ Cu has a smaller $N_{\rm{part}}$ and $N_{\rm{coll}}$. However,
one see that the background subtracted distributions in both
systems are qualitatively similar to each
other.

\vfill\eject
\end{document}